\documentclass[a4paper]{article}

\usepackage{INTERSPEECH2018}
\usepackage{tikz}
\usepackage[]{algorithm2e}

\newcommand{\bs}[1]{\boldsymbol{#1}}

\usetikzlibrary{shapes.misc}

\tikzstyle{module}=[shape=rounded rectangle, draw=black, minimum height=1cm]
\tikzstyle{data}=[shape=rectangle, fill=gray!50]

\newcommand\f{0.004cm}
\newcommand\cd{0.02cm}

\title{Capsule Networks for Low Resource Spoken Language Understanding}
\name{Vincent Renkens, Hugo Van hamme}
\address{Department Electrical Engineering-ESAT, KULeuven\\
Kasteelpark Arenberg 10, Bus 2441, B-3001 Leuven Belgium}
\email{vincent.renkens@esat.kuleuven.be, hugo.vanhamme@esat.kuleuven.be}

\begin{document}
\maketitle
\begin{abstract}

Designing a spoken language understanding system for command-and-control applications can be challenging because of a wide variety of domains and users or because of a lack of training data. In this paper we discuss a system that learns from scratch from user demonstrations. This method has the advantage that the same system can be used for many domains and users without modifications and that no training data is required prior to deployment. The user is required to train the system, so for a user friendly experience it is crucial to minimize the required amount of data. In this paper we investigate whether a capsule network can make efficient use of the limited amount of available training data. We compare the proposed model to an approach based on Non-negative Matrix Factorisation which is the state-of-the-art in this setting and another deep learning approach that was recently introduced for end-to-end spoken language understanding. We show that the proposed model outperforms the baseline models for three command-and-control applications: controlling a small robot, a vocally guided card game and a home automation task.

\end{abstract}
\noindent\textbf{Index Terms}: Spoken Language Understanding, Capsule Networks, Deep Learning, Low Resource

\section{Introduction}
\label{sec:introduction}

In this paper we will discuss a spoken language understanding (SLU) system for command-and-control applications. The system  can learn to map a spoken command to a task description. This description can then be given to some agent that can execute the task. An example for a command in a home automation application would be \textit{``Turn on the light in the kitchen''}. This could then be mapped to the task \texttt{Switch(kitchen light, on)}. The task is represented by the type of action, \texttt{Switch} in this example, and a collection of arguments, \texttt{kitchen light} and \texttt{on} in this example. The SLU system learns to map the spoken command to a semantic representation, which is a collection of labels, one for the action type and one for each of the arguments.\\
\\
Many approaches to this problem consist of an Automatic Speech Recognition (ASR) component that transforms the spoken command into a textual transcription and a Spoken Language Understanding (SLU) component that maps the textual transcription to the semantic representation \cite{wang2005spoken, de2008spoken}. Such a system makes some assumptions about the user and how they are going to use the system. The ASR is typically trained for a single language, so it is assumed that the user will use this language. If the user has a pronounced accent or if the user has a speech impairment the ASR will often introduce a lot of errors \cite{christensen2012comparative}, which makes it difficult for the SLU component to correctly determine the task to be performed. This is especially difficult for speech impaired users whose speech impairment is caused by another cognitive or motor disability. Users with such a disability often have difficulties using devices, so speech has a large potential to improve their way of living.\\
\\
Simple SLU components, like one based on key phrases assume that the user is going to use some predefined commands. From a design perspective choosing these key phrases can be difficult to impossible for some applications. More advanced methods based on Recurrent Neural Networks (RNN) \cite{dinarelli2016improving} or Conditional Random Fields \cite{raymond2007generative} need lots of data to train, which may not be available in the domain of the application.\\
\\
As an alternative we propose a system that learns to understand spoken commands directly from the user through demonstrations. The user can train the system by giving a spoken command and subsequently demonstrating the corresponding task through an alternative interface to the agent. The command \textit{``Turn on the light in the kitchen''} can be demonstrated by pressing the button to turn on the light. This demonstration can then be converted to a semantic representation. The system directly maps speech to the semantic representation, without going through an intermediate textual representation. The system is trained using only the data from the user, which means that the assumptions and restrictions mentioned above do not apply. The disadvantage of such a system is that the user needs to give some examples, which requires some effort on their part. In order to minimize this effort it is crucial that the required amount of training data is as small as possible.\\
\\
In the past we have proposed a method based on Non-Negative Matrix Factorisation (NMF) for this task \cite{gemmeke2013self, ons2014self}. NMF performs significantly better than other, more conventional approaches like Hidden Markov Model (HMM) based approaches \cite{gemmeke2014dysarthric}. There are however limitations to the NMF approach. For example, the NMF approach uses a bag-of-words representation. It does not consider the order in which the words occur, which can be important to correctly interpret the command \cite{renkens2017weakly}. Deep learning based approaches have shown great performance on many speech-related tasks \cite{hinton2012deep, chorowski2015attention, serdyuk2018towards}. These models are based on Deep Neural Networks or RNNs and require a lot of data to train, which is not available in this setting. In this paper we propose to use a capsule network with a bidirectional RNN encoder. Capsule networks were proposed in \cite{sabour2017dynamic} and it is suggested that they make more efficient use of the training data, making them better suited for this task.\\
\\
We will discuss our proposed model in section \ref{sec:model}. In section \ref{sec:experiments} we will describe the performed experiments and we will evaluate the results in section \ref{sec:results}. Finally we will end with some conclusions in section \ref{sec:conclusions}.

\section{Model}
\label{sec:model}

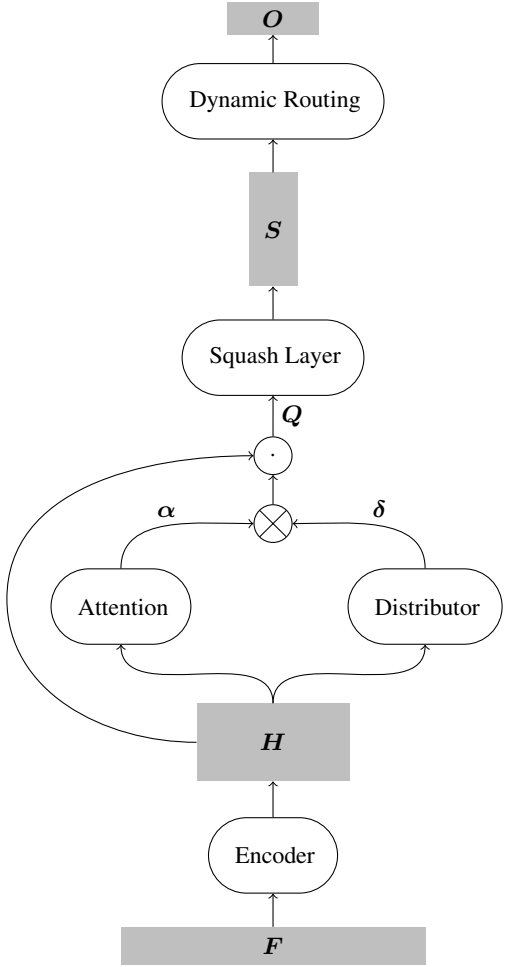
\begin{figure}[t]
	\centering
	\begin{tikzpicture}[
		cross/.style={path picture={\draw[black]
		(path picture bounding box.south east) -- (path picture bounding box.north west) (path 				picture bounding box.south west) -- (path picture bounding box.north east);
		}}]
		
		\node[data, minimum width=4cm, minimum height=123*\f] (features) at (0, 0) {$\bs{F}$};
		\node[module] (encoder) at (0, 1.2) {\hspace*{0.15cm} Encoder \hspace*{0.1cm}};
		\node[data, minimum width=2cm, minimum height=256*\f] (encoded) at (0, 2.7) {$\bs{H}$};
		\node[module] (attention) at (-2, 4.5) {\hspace*{0.15cm} Attention \hspace*{0.1cm}};
		\node[module] (distributor) at (2, 4.5) {\hspace*{0.15cm} Distributor \hspace*{0.1cm}};
		\node [draw, circle, cross, minimum width=0.5cm] (multiply) at (0, 5.6) {}; 
		\node [draw, circle, minimum width=0.5cm] (dot) at (0,6.5) {$\cdot$}; 
		\node[module] (squash) at (0, 7.8) {\hspace*{0.15cm} Squash Layer \hspace*{0.1cm}};
		\node[data, minimum width=32*\cd, minimum height=75*\cd] (capsules) at (0, 9.5) {$\bs{S}$};
		\node[module] (dynamic) at (0, 11.2) {\hspace*{0.15cm} Dynamic Routing \hspace*{0.1cm}};
		\node[data, minimum width=60*\cd, minimum height=16*\cd] (output) at (0, 12.3) {$\bs{O}$};
		\draw [->] (features) -- (encoder);
		\draw [->] (encoder) -- (encoded);
		\draw [->] (encoded) to [out=90, in=270] (attention.south);
		\draw [->] (encoded) to [out=90, in=270] (distributor.south);
		\draw [->, out=90, in=180] (attention.north) to node[above]{$\bs{\alpha}$} (multiply);
		\draw [->, out=90, in=0] (distributor.north) to node[above]{$\bs{\delta}$} (multiply);
		\draw [->] (multiply) -- (dot);
		\draw [-, out=180, in=270] (encoded.west) to (-3.5, 4.6);
		\draw [->, out=90, in=180] (-3.5, 4.6) to (dot);
		\draw [->] (dot) -- (squash) node[midway, right]{$\bs{Q}$};
		\draw [->] (squash) -- (capsules);
		\draw [->] (capsules) -- (dynamic);
		\draw [->] (dynamic) -- (output);
	\end{tikzpicture}
	\caption{A schematic of the proposed model}
	\label{fig:schematic}
\end{figure}

\noindent Our proposed model is presented in figure \ref{fig:schematic}. The inputs to the model are a sequence of filter bank features $\bs{F}$. The inputs are first encoded into high level features $\bs{H}$. For this work a multi-layered bidirectional GRU \cite{cho2014learning} is used for this purpose. The sequence is Sub-sampled with a stride of 2 between the layers of the encoder as proposed in \cite{bahdanau2016end}. The sequence of high level features is therefore shorter than the sequence of input features.\\
\\
The sequence of high level features is converted into capsules $\bs{S}$ using an attention mechanism \cite{bahdanau2014neural} and a distributor. The concept of capsules was proposed in \cite{sabour2017dynamic}. A capsule is represented by a vector. The direction of the vector represents the latent properties of the capsules. The norm of the vector lies between 0 and 1 and represents a probability that the capsule is present or not.\\
\\
The attention module is used to determine a weight representing the importance of each timestep. Not the entire sequence is important to determine the meaning of the utterance (e.g. words like ``please''). The attention module gives the model the capability to filter out the unimportant parts. The attention weights are determined using a sigmoid layer with a single output on the high level features:

\begin{equation}
	\alpha_t = \text{sigmoid}(\bs{w}_\text{a} \cdot \bs{h}_t + b_\text{a})
\end{equation}

\noindent $\alpha_t$ is the attention weight for time $t$, $\bs{w}_\text{a}$ and $b_\text{a}$ are the weights and bias of the sigmoid layer and $\bs{h}_t$ contains the high level features for time $t$.\\
\\
The distributor is used to distribute each timestep to the hidden capsules $\bs{S}$. A distribution weight is determined from each timestep to each hidden capsule. Similar to the attention weights, the distribution weights are determined using a softmax layer on the high level features:

\begin{equation}
	\bs{\delta}_t = \text{softmax}(\bs{W}^\text{d} \cdot \bs{h}_t + \bs{b}^\text{d})
\end{equation}

\noindent $\bs{\delta}_t$ contains the distribution weights for timestep $t$, one for each hidden capsule. $\bs{W}^\text{d}$ and $\bs{b}^\text{d}$ are the weights and biases of the softmax layer. Using the attention and distribution weights a context vector is created for each hidden capsule $i$:

\begin{equation}
	\bs{q}_i = \sum_t \alpha_t\delta_{ti}\bs{h}_t
\end{equation}

\noindent Where $\bs{q}_i$ is the context vector for capsule $i$. The context vectors are then converted to the capsule representation using a squash layer. The squash layer is a linear transformation followed by a squashing function:

\begin{equation}
	\bs{s}_i = \sigma(\bs{W}^\text{s} \cdot \bs{q}_i)
\end{equation}

\noindent $\bs{s}_i$ is the vector representation for capsule $i$, $\bs{W}^\text{s}$ are the weights of the squashing layer. Notice that no bias is included in the linear transformation to ensure that context vectors with a small norm result in capsules with a small norm. $\sigma(\cdot)$ is the squashing function as defined in \cite{sabour2017dynamic}:

\begin{equation}
	\sigma(\bs{x}) = \frac{||\bs{x}||^2}{1+||\bs{x}||^2}\frac{\bs{x}}{||\bs{x}||}
\end{equation}

\noindent The squashing function ensures that the norm of the vector representations lies between 0 and 1. The output capsules $\bs{O}$ are computed using the iterative dynamic routing algorithm proposed in \cite{sabour2017dynamic}. Every hidden capsule will predict the output of every output capsule using a linear transformation:

\begin{equation}
	\bs{p}_{ij} = \bs{W}^\text{p}_{ij}\cdot\bs{s}_i
\end{equation}

\noindent $\bs{p}_{ij}$ is the predicted vector representation of output capsule $j$ from hidden capsule $i$ and $\bs{W}^\text{p}_{ij}$ contains the weights for this prediction. The output capsules are computed using the coupling coefficients $\bs{C}$. The coupling coefficients represent how strongly linked the hidden capsules are to the output capsules. The coupling coefficients are computed using a softmax function on the coupling logits $\bs{B}$. The coupling logits are initialised with learnable values and then iteratively fine tuned with the dynamic routing algorithm:\\

\begin{algorithm}[h]
	Define variable $\bs{B}^{(1)}$\;
	\For{$n=1$:$N$ }{
		For all hidden capsules $i$: $\bs{c}_i = \text{softmax}(\bs{b}^{(n)}_i)$\;
		For all output capsules $j$: $\bs{o}_j = \sigma(\sum_i c_{ij}\bs{p}_{ij})$\;
		For all logits $b^{(n)}_{ij}$ in $\bs{B}^{(n)}$: $b^{(n+1)}_{ij}$ = $b^{(n)}_{ij} + \bs{p}_{ij}\cdot\bs{o}_j$\;
	}
	\caption{Dynamic routing algorithm}
\end{algorithm}

\noindent At each iteration the output capsules are computed with the current connection logits. The connection logits are updated based on the agreement between the output capsule and the prediction. The agreement is measured using the scalar product. If the agreement between a prediction from a hidden capsule with an output capsule is large the connection logit will increase. The dynamic routing algorithm will look for groups of similar predictions for each output capsule. If there is a group of predictions that agree for a certain output capsule the capsule will become active and its norm will be close to one. If there is no such group the capsule will be inactive and its norm close to zero.\\
\\
The probabilities of the output labels $\bs{l}$ are finally computed using the norm of the output capsules:

\begin{equation}
	l_j = ||\bs{o}_j||
\end{equation}

\noindent The network is trained by minimizing the margin loss:

\begin{equation}
	L = \sum_j t_j\max(0, 0.9 - l_j) + (1-t_j)\max(0, l_j - 0.1) 
\end{equation}

\noindent where $t_j$ is the target for label $j$, which is either 0 or 1.

\section{Experiments}
\label{sec:experiments}

\subsection{Datasets}

The proposed model is tested and compared to the baselines for three datasets, in the domains of robotics, a card game and home automation.\\
\\
The \texttt{GRABO} \cite{renkens2014acquisition} dataset contains English and Dutch commands given to a robot. The robot can move in its environment, pick up objects and use a laser pointer. Typical commands given to the robot are ``\textit{move to position x}'' or ``\textit{grab object y}''. Output labels include positions in the world, the actions the robot can take etc. There are a total of 30 output labels. Data was recorded from 11 speakers issuing 36 different commands with 15 repetitions.\\
\\
The \texttt{PATCOR} dataset \cite{tessema2013metadata} contains Dutch utterances from a vocally guided card game called Patience. The players can move cards or get new cards from the deck. Typical commands are ``\textit{Put card x on card y}'' or ``\textit{New cards}''. The output labels are the value and suit of the card being moved, the target position etc. There are a total of 38 output labels. Data was recorded from 8 speakers.\\
\\
The \texttt{DOMOTICA-3} dataset \cite{ons2014self} is a follow up of the \texttt{DOMOTICA-2} dataset \cite{tessema2013metadata} and contains utterances from Dutch dysarthric speakers using voice commands in a home automation task. Typical commands are ``\textit{open door x}'' or ``\textit{turn on light y}''. The output labels include all the lights, doors and all the actions the system can take. There is a total of 25 output labels. Data was recorded from 17 speakers with varying levels of dysarthria. Because speaking costs more effort for some speakers the amount of data per speaker varies greatly.

\subsection{Methodology}

We use cross-validation to get reliable experimental result. First, we split the data in multiple blocks. The blocks are chosen such that they are maximally semantically similar. This is done by minimizing the Jensen-Shannon Divergence between the blocks. We then create the training set by taking the data from a random set of blocks and the rest of the data is put in the testing set. We create learning curves by putting an increasing number of blocks in the training set. To get more reliable results we do 5 experiments for each number of blocks in the training set, each time with a different set of random blocks.\\
\\
40 Mel filter banks + energy including first and second order derivatives with a window size of 25 ms and a window step of 10 ms are used as input features. A voice activity detector is used to remove long silences from the commands. The encoder consists of 2 bi-directional GRU layers with 256 units. There are 32 hidden capsules in $\bs{S}$ with 64 dimensions. There is one output capsule with a dimension of 8 for every output label. In total the network has around 2.2 million parameters. The model is trained with batches of 16 utterances for 30 epochs. Adam \cite{kingma2014adam} is used as the optimization method with a learning rate of 0.001.

\subsection{Baseline}

As a first baseline we use the method proposed in \cite{gemmeke2013self, ons2014self}. This method is based on Non-negative Matrix Factorisation (NMF) that is used to decompose the input utterance into recurring patterns, which can be thought of as words. These words are linked to the output labels and in such a way a dictionary of words corresponding to the labels is created. This method achieves state-of-the-art performance for this task \cite{gemmeke2014dysarthric}.\\
\\
Alternatively we use a different deep learning approach proposed in \cite{serdyuk2018towards} as a second baseline. This model was proposed in the context of end-to-end NLU to predict domain and intent labels for spoken utterances. The model consists of the same encoder, with the same number of layers and units, used in the current paper and a decoder. The decoder aggregates the high level features with max-pooling then applies a hidden ReLU layer with 1024 units followed by a sigmoid output layer to get the probabilities of the output labels. This network has around 2.3 million parameters. Adding more layers to the encoder did not improve the results. We will refer to this model as ``encoder-decoder'' in the results section.

\section{Results}
\label{sec:results}

\begin{figure}[ht!]
\centering
\vspace*{0.7cm}
\includegraphics[width=1\linewidth]{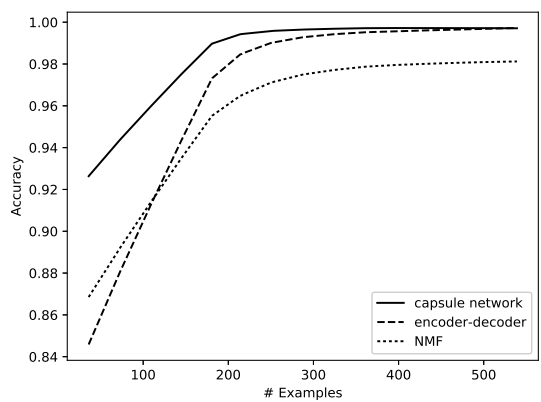}
(a)
\includegraphics[width=1\linewidth]{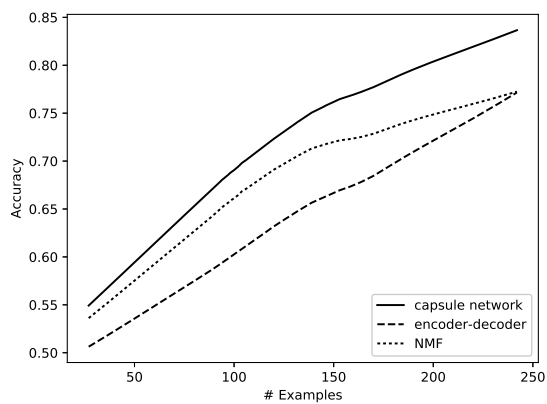}
(b)
\includegraphics[width=1\linewidth]{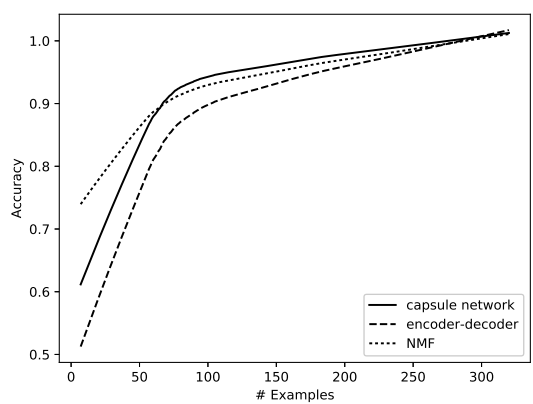}
(c)
\caption{The accuracy plotted in function of the number of examples obtained by the proposed system and the baseline systems for the \texttt{GRABO} dataset (a), the \texttt{PATCOR} dataset (b) and the \texttt{DOMOTICA-3} dataset (c). LOWESS smoothing is used to obtain a smooth curve.}
\label{fig:results}
\vspace*{0.3cm}
\end{figure}

In figure \ref{fig:results} the accuracy of the models is plotted as a function of the number of examples in the training set for all three data sets. In most cases the proposed model outperforms the baseline models. Only for the \texttt{DOMOTICA-3} dataset, for a very small training set, the NMF model outperforms the capsule network. This may be caused by the fact that \texttt{DOMOTICA-3} contains dysarthric speech, which is less consistent in terms of timing and pronunciation. NMF does not suffer a lot from this variability, but the GRU encoder might have more trouble modelling it. However, with a little bit more data the capsule network catches up with NMF and performs slightly better. For the \texttt{GRABO} dataset all models achieve a high accuracy, but the capsule network performs best. The encoder-decoder model does not perform well for small amounts of data, but catches up if more data is available.\\
\\
For the \texttt{PATCOR} dataset the accuracy for all models is significantly lower. This is because \texttt{PATCOR} is a more challenging dataset. The user can look at the state of the game and they might leave out information in the command because it is obvious from the state of the game. For example if there is only one $3$ that can be moved they might not mention the suit of the card to be moved. The state of the game is however not available to the NLU system in this setup, which introduces errors. In Dutch there are several names for each card. Some users alternate between these names, which also makes it more challenging for the NLU. Even on this more challenging task the capsule network performs better than the NMF model, especially with more training data. The encoder-decoder seems to have trouble with this more challenging task, which supports the findings in \cite{vukotic2015time}\\
\\
It is remarkable that the capsule network performs so well for only a couple dozen examples, which amounts to a few minutes of speech. These experiments seem to support the hypothesis that capsule networks make more efficient use of the training data, especially when you compare the capsule network with the encoder-decoder for small amounts of data. 

\section{Conclusions}
\label{sec:conclusions}

In this paper we proposed a capsule network for low resource spoken language understanding for command-and-control applications. Only the data from the user is used to train the system, making it able to adapt to the domain of the application and the speaker without needing training data prior to deployment. The proposed model has been shown to significantly outperform the previous state-of-the-art. Even for small amounts data, a few dozen utterances, the capsule network performs well. In future work we will look more closely at the reason why the capsule network works well, especially for so little training data. It might also be interesting to investigate using a distributer together with an attention mechanism for attention based speech recognition.

\section{Acknowledgements}

The Research in this work was funded by PhD grant 151014 of the Research Foundation Flanders (FWO)

\bibliographystyle{IEEEtran}

\bibliography{mybib}

\end{document}